\documentclass[a4paper,pre,superscriptaddress,twocolumn,amsfonts,amsmath,floatfix]{revtex4-1}
\usepackage[sort&compress]{natbib}
\usepackage{color}      
\usepackage{epsfig}
\newlength\figurewidth
\AtBeginDocument{\setlength\figurewidth{.5\linewidth}}

\def\kT{\ensuremath{k_\text{B}T}}
\def\Fext{\ensuremath{F_\text{ext}}}
\def\gammaeff{\ensuremath{\gamma_\text{eff}}}

\def\url#1{}

\begin{document}
\title{Dynamics and friction of a large colloidal particle in a bath of hard spheres: Langevin dynamics simulations and hydrodynamic description}
\date{\today}
\def\unikn{\affiliation{%
  Fachbereich Physik, Universit\"at Konstanz,
  78457 Konstanz, Germany}}
\def\unial{\affiliation{%
  Departamento de F\'\i{}sica Aplicada, Universidad de Almer\'\i{}a,
  04.120 Almer\'\i{}a, Spain}}
\def\unialinfo{\affiliation{%
Departamento de Inform\'atica, Campus de Excelencia Internacional Agroalimentario (ceiA3), Universidad de Almer\'\i{}a,
  04.120 Almer\'\i{}a, Spain}}
\author{F. Orts}\unialinfo
\author{G. Ortega}\unialinfo
\author{E.M. Garz\'on}\unialinfo
\author{M.~Fuchs}\unikn
\author{A.~M.~Puertas}\unial

\begin{abstract}
The analysis of the dynamics of tracer particles in a complex bath can provide valuable information about the microscopic behaviour of the bath. In this work, we study the dynamics of a forced tracer in a colloidal bath by means of Langevin dynamics simulations and a theory model within continuum mechanics. In the simulations, the bath is comprised by quasi-hard spheres with a volume fraction of $50\%$ immersed in a featureless quiescent solvent, and the tracer is pulled with a constant small force (within the linear regime). The theoretical analysis is based on the Navier Stokes equation, where a term proportional to the velocity arises from coarse-graining the friction of the colloidal particles with the solvent. As a result, the final equation is similar to the Brinkman model, although the interpretation is different. A length scale appears in the model, $k_0^{-1}$, where the transverse momentum transport crosses over to friction with the solvent. The effective friction coefficient experienced by the tracer grows with the tracer size faster than the prediction from Stokes law. Additionally, the velocity profiles in the bath decay faster than in a Newtonian fluid. The comparison between simulations and theory points to a boundary condition of effective partial slip at the tracer surface. We also study the fluctuations in the tracer position, showing that it reaches diffusion at long times, with a subdiffusive regime at intermediate times. The diffusion coefficient, obtained from the long-time slope of the mean squared displacement, fulfills the Stokes-Einstein relation with the friction coefficient calculated from the steady tracer velocity, confirming the validity of the linear response formalism.

\end{abstract}

\maketitle

\section{Introduction}
\label{sect_intro}

In soft matter, different time and length scales are involved, due to the presence, typically, of simple solvents and macromolecules. This is usually tackled by integrating out the fastest degrees of freedom, which leaves an equation of motion for the relevant (macromolecular) ones \cite{Chaikin1995,Dhont1996,NievesPuertas2016}. A clear example is the Langevin equation for the Brownian motion of a colloidal particle, where the solvent is modelled only through the friction and random forces acting on the particle. This allows the calculation of parameters characterizing the solvent by studying the diffusion of a single particle. This idea has been elaborated further to study more complex fluids, and is the core of so-called microrheology. 

In microrheology, a single colloidal tracer (or a very small number of them) is introduced in a complex fluid to study its mechanical behaviour at the microscopic scale \cite{Mason1995,Cicuta2007,Wilson2011-review, Puertas2014,Furst2017}. The tracer can be left undisturbed, to undergo diffusion in the complex bath due to thermal and density fluctuations (passive microrheology) or forced to probe the response of the bath (active microrheology). Experiments \cite{Habdas2004,Sriram2009,Sriram2010} and simulations of active microrheology \cite{Carpen2005,Khair2006,Gazuz2009,Zia2012,Winter2012,Zia2018} have shown that the effective friction coefficient shows a linear dependence on the force for small forces, allowing the definition of a microviscosity. A non-linear regime is entered for larger forces and a second linear regime, featured by a smaller viscosity may be attained for large forces. This overall phenomenology resembles that of conventional (bulk) rheology, showing shear thinning, thickening or more complex scenarios \cite{Larson1999,Brady1993,Strating1999}. Different possibilites have also been reported in microrheology, depending on the interactions considered \cite{Zia2018}.

The interpretation of the results from microrheology must take into account all the degrees of freedom. While in dilute cases, theory achieves to consider the bath particles explicitly (e.g. by the direct interactions between the tracer and bath particles, or among the bath particles) \cite{Squires2005,Squires2008,DePuit2011,Leitmann2013,Puertas2014,Leitmann2017, Leitmann2018}, a dense fluid is often described within hydrodynamics. This implies that not only the solvent, but also the bath must be treated as continuum fluid \cite{Larson1999}. While the solvent is typically a Newtonian fluid, the bath is a complex one, namely, the transport coefficients depend on the driving. The models used in microrheology, thus, must describe the interaction of the tracer with these two baths, either as fluids with different properties \cite{Levine2000,Levine2001,Felderhof2009,Chu2019}, sacrificing a detailed structural description, or using a microscopic theory to describe the motion of the tracer and bath particles in a solvent \cite{Gazuz2013,Gruber2016,Gruber2020}. 

In this work, we study the dynamics of a large tracer in a dense bath of colloidal particles; the tracer is subjected to an external constant force, small enough to remain in the linear regime. 
All particles exhibit Langevin motion characterized by a constant friction coefficient with the solvent, which also provides the random forces. They are taken to be Gaussian and white, and the fluctuation dissipation relation holds. This widely used model 
focuses on the collective interactions among bath particles and tracer, while it neglects the solvent flow, which leads to hydrodynamic interactions \cite{Dhont1996}.
We have run simulations with a tracer up to eight times larger than the bath particles, and a bath volume fraction of $\phi=0.50$. The results are analysed using a hydrodynamic model, within the formalism of continuum mechanics. It differs from the (naively expected) Navier Stokes hydrodynamics even in the limit of macroscopic tracers. The model has been derived coarse-graining systems of Langevin particles \cite{Vogel2019}, and the resulting hydrodynamic equation is the Brinkman equation, which has been applied previously to diffusion in porous systems \cite{Brinkman1947}, although our interpretation is different from previous ones. Notably, the solution of the Brinkman equation brings out a length scale where  transverse momentum transport crosses over to friction with the solvent. The friction coefficient thus grows with the tracer size much faster than the Stokes' law while the velocity profile in the bath decays as the inverse cubed distance to the tracer. After performing a finite size analysis in the simulation results, the friction coefficient and velocity profile can be correctly rationalized within the theoretical model. The effect of the different boundary conditions on the tracer surface is also discussed. Finally, we study the dynamics of the tracer using the mean squared displacement and confirm the validity of the Stokes-Einstein relation for all tracer sizes.

\section{Model}
\label{model}

The system we aim to describe is a colloidal bath at high density with a (colloidal) tracer particle equal or larger than the bath particles. There are, therefore, three components in the system: solvent, bath particles, and tracer particle. While the system is in equilibrium, at time $t=0$ a constant external force starts to pull the tracer. Similar systems have been considered to study microrheology both in simulations \cite{Winter2012, Schroer2013, Gradenigo2016,Gruber2016,Wang2016,Leitmann2017} and in theory \cite{Squires2005,Gazuz2009,DePuit2011,Wulfert2017}. In our case, the force is small enough to drive the system out of equilibrium within the linear regime. 

We approach this system from two points of view: using Langevin dynamics simulations and a theoretical model based on continuum mechanics. In both cases, the solvent is assumed to be at rest, its only effect being a friction force proportional to the particles velocities, and a random force which produces Brownian motion. This implies that we neglect hydrodynamic interactions (HI) among all particles, but allows us to run simulations of large systems and proceed analytically in the theory, and connects with many previous works where HI are also neglected. This may seem a harsh approximation but its effect on the local cageing of particles is only quantitative \cite{Zia2018,Marenne2017}, not affecting the physical behaviour of the system, in particular at the high bath density studied here.

\subsection{Simulations}
\label{sect_sim}

In the simulations, the system under study is composed of $N$ polydisperse particles, including a tracer (labeled with $j=1$), in a cubic box with periodic boundary conditions. All particles undergo Brownian motion, which we model by the Langevin equation \cite{Dhont1996}. For particle $j$, the equation of motion reads:

\begin{equation}
m_j \frac{d^2\, {\bf r}_j}{dt^2}\:=\: \sum_{i\neq j} {\bf F}_{ij} - \gamma_j \frac{d\, {\bf r}_j}{dt} + {\bf f}_j(t) + {\bf F}_{ext} \delta_{j1}
\label{Langevin}
\end{equation}

\noindent where $m_j$ is the particle mass, ${\bf F}_{ij}$ is the interaction force between particles $i$ and $j$, $\gamma_j$ is the friction coefficient with the solvent, assumed to be proportional to the particle radius $a_j$, $\gamma_j=\gamma_0 a_j$, mimicking Stokes' law, and ${\bf f}_j$ is the Brownian force. The latter is random, but its intensity is linked to the friction force, as given by the fluctuation-dissipation theorem, $\langle {\bf f}_j(t) \cdot {\bf f}_j(t') \rangle = 6 k_BT \gamma_j \delta(t-t')$, where $\kT$ is the thermal energy and $\delta(x)$ is the Dirac-delta symbol \cite{Dhont1996}. Finally, the external force, ${\bf F}_{ext}$, acts only on the tracer (as shown by the Kronecker-delta symbol, $\delta_{j1}$). The energy injected by this force is dissipated by the friction of the tracer with the solvent and the bath particles, keeping the kinetic temperature constant in the stationary state. As mentioned above, hydrodynamic interactions have been neglected in the equation of motion. 

The interaction potential between particles $i$ and $j$ is derived from the central inverse-power potential:

\begin{equation}
V({\bf r})\:=\:\kT \left( \frac{r}{a_{ij}} \right)^{-36} \label{potential}
\end{equation}

\noindent with $r=\left| {\bf r} \right|$ the center to center distance between the particles and $a_{ij}=a_i+a_j$. Due to the high value of the exponent, this system behaves as colloidal hard spheres \cite{Lange2009}. To avoid crystallization at high density, a continuous size distribution of width $2\delta=0.2a$, with $a$ the mean radius, is used for the bath particles. The tracer has radius $a_t\geq a$. For the sake of simplicity in the numerical algorithm, all particles, including the tracer, have the same mass: $m_j=m$ (note that the tracer particle gives a scale for the external force). The mean bath particle radius, $a$, the thermal energy $k_BT$, and particle mass $m$ are the length, energy and mass units, respectively. The friction coefficient with the solvent of particle $j$ is calculated with $\gamma_0=5\,\sqrt{m \kT}/a$, which gives a single particle diffusion coefficient of $D_0 = \kT/\gamma_0 = 0.2\,a \sqrt{\kT/m}$ for the mean particle. The Langevin equations of motion are integrated using the Heun algorithm \cite{Paul1995}, with a time step of $\delta t=0.0005 \, a \sqrt{m/\kT}$.  

In our simulations, the system containing the tracer is equilibrated without external force. For $t>0$, the constant external force is applied onto the tracer in the $z$-direction, and its trajectory is monitored. The effective friction coefficient experienced by the tracer is obtained from its long time steady velocity, $\langle v \rangle$, averaged over many independent trajectories, and using the steady-state relationship $\Fext = \gammaeff \langle v \rangle$. For small forces, the tracer velocity presents a linear regime with the external force, resulting in a constant friction coefficient, followed by a decrease of $\gammaeff$ for larger forces (non-linear response) \cite{Orts2019}. We focus here on the linear regime at small forces. The connection to the hydrodynamic calculation of the theoretical section below is then given by Onsager's regression hypothesis \cite{Furst2017}. 

\begin{figure}
\includegraphics[width=0.45\textwidth]{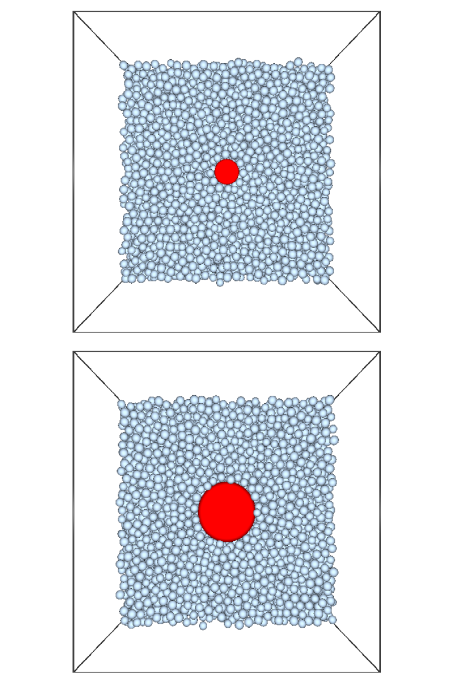}
\caption{Snapshots of the systems with $N=15625$, with a tracer with $a_t=3a$ (top panel) and $a_t=7a$ (bottom panel). The tracer is marked in red, and the particles in front of it have been removed for clarity. \label{snapshots}}
\end{figure}

In hydrodynamics, it is well known that there are long range correlations in the fluid, decaying typically with the inverse distance. Although our model predicts a faster decay of these correlations, as shown below, it is mandatory to perform an analysis of finite size effects. In fact, since periodic boundary conditions are used, an infinite cubic array dragged through a bath of particles is considered. We have thus run simulations of systems with $N=216$, $512$, $1000$, $2197$, $4096$, $8000$, $15625$ and $32768$ particles, and tracer sizes from $a_t=a$ to $a_t=8a$. Fig. \ref{snapshots} presents two snapshots of the system with $N=15625$ particles, including a tracer of size $a_t=3a$ (top panel) and $a_t=7a$ (bottom panel). The bath volume fraction is $\phi=0.50$ in all cases, and the volume occupied by the tracer is not accounted for in the calculation of the simulation box size \cite{Orts2019}. The center of mass of the system is not fixed when the external force is applied. For the calculation of the friction coefficient, $500$ tracer trajectories have been analysed. In addition to the tracer dynamics, the density and velocity profile in the bath has been studied in several cases to check the theoretical predictions; note that Langevin dynamics gives directly the particle velocity.

\subsubsection{Numerical implementation}

From a computational point of view, the requirement of a finite size analysis implies running simulations with different number of particles, $N$, for every tracer size, $a_t$. For this purpose we have used high performance computing in two ways: \textit{i}) programming in Graphics Processing Units (GPU) to speed up the simulation of a single trajectory, and \textit{ii}) using a Genetic Algorithm (GA) to balance the load of all the processing units of the computer cluster, taking into account the different durations of the simulations with different $N$.

We have accelerated the computation of a single tracer trajectory by means of GPU computing using the CUDA interface~\cite{Orts2019,Ortega2017,ICA3PPOrtega2016}. Note that the full system with $N$ particles has to be simulated, although the trajectory of a single particle (the tracer) is the most relevant. In particular, the calculation of the interaction forces among all particles and the integration of the equations of motion are very demanding, and have been thoroughly optimized \cite{Morozov2011}. In addition to this CUDA-GPU core, a standard sequencial FORTRAN code has been used in the CPUs. It was checked that both codes give the same results when the same sequence of random numbers is used for the Brownian force. 

The whole set of simulations to analyze the friction coefficient has been run on modern Multi-GPU clusters, that provide CPU-cores and GPUs which can compute several simulations in parallel. Since the simulations of systems with different sizes are needed, the computational loads of the corresponding tasks are also different. Therefore, an appropriate balance for the execution is decisive. Here, we have adapted a genetic algorithm to achive the optimal parallel performance \cite{Sels2015}. In our GA, a set of possible solutions of the scheduling problem is the \textit{population}. The algorithm evolves iteratively, starting with a random population, using the mutation and selection mechanisms until the optimal solution is reached, as defined by the minimum spread in execution times among all processing units. A code written in Python has been developed to calculate the optimal distribution of tasks. 

In our procedure, a single trajectory in every unit (CPU-core or GPU-core) is executed for every size and a given tracer radius, and the running times are recorded. With these times, the optimal distribution of trajectories per unit is calculated, ensuring that all units finish their tasks with a minimum difference. This distribution is then passed to the cluster to perform the whole set of simulations for a single tracer size. As mentioned above, simulations with $N=216, 512, 1000, 2197, 4096, 8000, 15625$ and $32768$ particles have been run, ensuring that all particles can fit into the simulation box (recall that the tracer volume is not accounted for in the calculation of the simulation box size). Thus, for large tracers, only the biggest systems are simulated. Every trajectory has been recorded for $10^6$ time steps, corresponding to $t=500\, a \sqrt{m/\kT}$, or $t=100 a^2/D_0$. This time is long enough to reach the stationary state and provide a correct estimation of the tracer velocity, as checked with longer simulations in selected cases.

A cluster composed by 4 nodes with a multiprocessor of 16 CPU-cores (Bullx R424-E3 Intel Xeon E5 2650 with 8GB RAM) and 2 GPUs NVIDIA Tesla M2070 has been used. Table~\ref{tab:CPUGPUtimes} shows the runtime  on a CPU-core and a GPU to simulate a single trajectory (profiling stage) for the systems with $a_t=3a$. Note that GPU-programming is particularly advantageous for large systems (up to $24\times$ faster), although the sequencial code is faster for small systems. 

\begin{table}[hbt!]
\centering

\begin{tabular}{r @{\hskip 0.2in}r @{\hskip 0.2in}r @{\hskip 0.2in}r }\hline
         $N$ & $t_{GPU}$ &   $t_{CPU} $ \\ \hline
       216 &    1580 &     790 \\
       512 &    1785 &    1860 \\
      1000 &    2240 &    3715 \\
      2197 &    2930 &    8710 \\
      4096 &    4450 &   18065 \\
      8000 &    7650 &   43080 \\
     15625 &   12050 &  113940 \\
     32768 &   20012 &  479313 \\      \hline
\end{tabular}
\caption{ Runtime in seconds of the simulation of a single trajectory for different sizes ($N$). $t_{GPU}$ and  $t_{CPU}$  columns show the execution time for a single trajectory on a GPU NVIDIA Tesla M2070/a and CPU-core Bullx R424-E3, respectively.}
\label{tab:CPUGPUtimes}
\end{table}

\subsection{Theory}
\label{theory}

We now search for a continuum mechanics description, in order to understand the motion of a macroscopic tracer in the bath of interacting Brownian particles.  
This search is motivated by the success of Stokes' calculation of the friction of a macroscopic tracer in a Newtonian fluid. He obtained it based on the Navier-Stokes equation (NSE) for the velocity of a continuous Newtonian fluid subjected to external stresses or forces. In colloid science, Stokes' law describes an isolated rigid  particle immersed in a solvent which is dragged with a constant velocity, with stick (or slip) boundary conditions on the particle surface. The resulting friction force depends linearly on the solvent viscosity and the bead radius, and is proportional to its velocity. 

Here, to describe the tracer in a colloidal bath we have to coarse-grain the system of coupled Langevin equations for the bath particles $j=2,\dots,N$ in Eq.~\eqref{Langevin}.  This was recently performed using the Zwanzig-Mori projection operator technique \cite{EvansMorriss,HessKlein} and considering the long-wavelength limit \cite{Vogel2019}.  The presence of the solvent leads to the inclusion of an additional friction term in the NSE, proportional to the bath particle velocity field, $\bf u$. This accounts for the local dissipation of the bath particles in the solvent, and arises from coarse-graining the drag forces on the particles  \cite{Vogel2019}. In the stationary state, the hydrodynamic equation reads:

\begin{equation}
{\bf \nabla} P - \eta_0 \nabla^2 {\bf u} = - \zeta_0 {\bf u} + {\bf F_{ext}} \label{brinkman_eq}
\end{equation}

\noindent where $P$ is the pressure. This equation contains the hydrodynamic friction with a bath of viscosity $\eta_0$ (that represents the colloidal system), and with an inert solvent, of friction coefficient $\zeta_0$, (representing the solvent) as well as an external force acting on the system. Without hydrodynamic interactions, the friction coefficient in incompressible systems is simply $\zeta_0=n\gamma_0$ where $n$ is the bath density. For the calculation of the analogue of Stokes' friction, the external force $\bf F_{ext}$ is assumed to be a point force acting on the tracer center. This equation is complemented by the incompressibility condition:

\begin{equation}
{\bf \nabla} \cdot {\bf u} = 0 \label{compress}
\end{equation}

Eq. (\ref{brinkman_eq}) was already proposed by Brinkman to describe the motion of a tracer in a swarm of colloidal particles \cite{Brinkman1947}, as a combination of Darcy's equation and the NSE. However, the interpretation of the parameters is different: in the Brinkman model, the divergence of the stress tensor represents the solvent, and the linear term in ${\bf u}$ is due to the presence of the other particles, which act as a porous matrix. Tam \cite{Tam1969} used a more rigorous derivation to this equation from first principles, albeit with the same interpretation. Due to this interpretation, the Brinkman equation has been widely used to study the diffusion in a porous medium \cite{Durlofsky1987}. It must be also mentioned that the Brinkman equation is similar to the Laplace-transformed unsteady Navier-Stokes equation.

It has been shown previously \cite{Vogel2019} that Eq.~(\ref{brinkman_eq}) holds with or without hydrodynamic interactions. It requires that momentum is not conserved (as holds in the Langevin simulations, where the solvent relaxes the momenta), yet that the bath viscosity $\eta_0$ is large in order for a region (later identified by the wavevector $k_0$) to emerge where the NSE holds in approximation. As any continuum mechanics description, application of  Eq.~(\ref{brinkman_eq}) requires smooth and slow fluctuations, which translates into large tracer sizes. As specific approximation,  Eq.~(\ref{brinkman_eq}) neglects the diffusive build-up of  a density profile around the forced tracer, which could become noticeable in an appreciably compressible system. It is also interesting to note  that the Brinkman's equation is not Galilei invariant, which is different from the NSE. This is in agreement with the Langevin equation, which is also not Galilei invariant. On the other hand, this implies that the problem of the moving sphere in a quiescent fluid is different from a fixed sphere in an incoming fluid. The problem we are interested in is the former one, namely, a moving tracer in a quiescent fluid.

This case has been solved previously in the literature, see e.g. \cite{Pozrikidis97}, giving a velocity profile around the tracer (located at ${\bf r}=0$):

\begin{equation}
{\bf u}({\bf r})=\frac{1}{8\pi \eta_0} {\cal S}({\bf r})\cdot {\bf F}_s + {\bf u}_{hom}({\bf r}) \label{u-brinkman}
\end{equation}

\noindent where ${\cal S}({\bf r})$ is a matrix of elements:

\begin{equation}
{\cal S}_{ij}({\bf r}) = \delta_{ij} \frac{{\cal A}(r)}{r} + \frac{r_ir_j}{r^3} {\cal B}(r)
\end{equation}

\noindent with

\begin{equation}
{\cal A}(r)=2\left( 1 + \frac{1}{k_0r} + \frac{1}{k_0^2r^2} \right) e^{-k_0r} - \frac{2}{k_0^2r^2}
\end{equation}

\begin{equation}
{\cal B}(r)=-2\left( 1 + \frac{3}{k_0r} + \frac{3}{k_0^2r^2} \right) e^{-k_0r} + \frac{6}{k_0^2r^2}
\end{equation}

\noindent and ${\bf F}_s=F_s \hat{e}_z$ is an effective surface force that depends on the boundary conditions (see below). The inverse distance $k_0$, appearing in the expressions above is defined as $k_0=\sqrt{\zeta_0/\eta_0}$ and describes the length scale of the crossover from friction at large distances, originating from the coupling of the particles to the solvent according to the Langevin equation, to diffusive transverse momentum transport intrinsic in the NSE based on Newtonian dynamics, for short distances. The ratio between this length scale and the tracer size, viz. the dimensionless parameter $k_0 a_t$, plays a central role in the following results; for $k_0\rightarrow 0$ the NSE description of a particle in a Newtonian solvent is recovered, whereas for $k_0 \rightarrow \infty$ the innert solvent is dominant. In particular, for small $k_0$:

\begin{equation}
\lim_{k_0\rightarrow 0} {\cal A}(r) = \lim_{k_0\rightarrow 0} {\cal B}(r) = 1
\end{equation}

\noindent what recovers the velocity profile for the Newtonian solvent \cite{Guyon}. 

The second term in eq. (\ref{u-brinkman}), ${\bf u}_{hom}$, is the velocity profile without external force and pressure, that decays exponentially:

\[{\bf u}_{hom} ({\bf r}) = - \hat{e}_r \frac{F_h a^2 e^{-k_0r}}{4\pi \eta_0r^3} \left(1+k_0r \right) \cos \theta \]
\begin{equation}
+ \hat{e}_{\theta} \frac{F_h a^2 e^{-k_0r}}{8\pi \eta_0r^3} \left(1+k_0r+k_0^2r^2 \right) \sin \theta
\end{equation}

\noindent Here, $F_h$ has to be determined by the boundary conditions, as well as $F_s$. For stick boundary conditions,

\[ {\bf u}(a_t)=u_0, \hspace{1cm} \mbox{and} \hspace{1cm} {\bf u}(r\rightarrow \infty)=0 \]

\noindent with $u_0$ the tracer velocity. This yields:

\[ F_s = 6 \pi \eta_0 a_t u_0 \left(1 + k_0 a_t + \frac{1}{3} k_0^2 a_t^2 \right) \]

\noindent and

\begin{equation}
F_h = -4 \pi \eta_0 a_t u_0 \left(1 + \frac{3}{k_0a_t}+\frac{3}{k_0^2 a_t^2}-3\frac{e^{k_0a_t}}{k_0^2a_t^2} \right)
\end{equation}

For slip boundary conditions, on the other hand, it is customary to introduce a slip length, $b$, and replace the condition of the surface velocity with 

\[ u_r(a_t)=u_0 \cos \theta, \hspace{0.6cm} \mbox{and} \hspace{0.6cm} \eta\left(u_{\theta}(a_t)+u_0\sin \theta \right) = b \tau_{r\theta} \]

\noindent where $u_r$ and $u_{\theta}$ refer to the radial and angular components of the velocity field, and $\tau_{r\theta}$ to the shear stress at the slip plane. For pure slip boundary conditions $b\rightarrow \infty$, resulting in \cite{Felderhof2007}:

\[ F_s = 6 \pi \eta_0 a_t u_0 \left[\frac{2 \left(1+k_0a_t\right) + k_0^2a_t^2 + k_0^3a_t^3/3}{3+k_0a_t} \right] \]

\noindent and

\begin{equation}
F_h = -4 \pi \eta_0 a_t u_0 \left[\frac{2\left(1+k_0a_t-e^{k_0a_t}\right)+k_0^2a_t^2+k_0^3a_t^3/3}{k_0^2a_t^2 \left(1+k_0a_t/3 \right)} \right]
\end{equation}

The friction force experienced by the tracer, equal to $\bf F_{ext}$, is calculated integrating the stress tensor over the tracer surface. For stick boundary conditions, this leads to \cite{Pozrikidis97}:

\begin{equation}
{\bf F_{ext}}=6 \pi \eta_0 a_t {\bf u_0} \left(1 + k_0 a_t + \frac{1}{9} k_0^2 a_t^2 \right) \label{fric_stick}
\end{equation}

\noindent Note that this expression reduces to Stokes formula for a Newtonian fluid, $\zeta=0$ (giving $k_0=0$), while in the opposite limit, $k_0 \rightarrow \infty$, or $\eta_0 \rightarrow 0$, the friction coefficient gives $V_t \zeta_0 /2$, with $V_t$ the volume of the tracer. 

The velocity profile from the Brinkman equation, Eq. (\ref{u-brinkman}), on the other hand, shows a faster decay than the NSE, as shown by the $\sim 1/(k_0^2 r^3)$ dependence at long distances. As expected, for $k_0=0$, the $~1/r$ decay, typical of a Newtonian fluid within the NSE, is recovered.

For slip boundary conditions the friction coefficient is given by:

\begin{equation}
{\bf F_{ext}}=6 \pi \eta_0 a_t {\bf u_0} \left(\frac{2 + 2 k_0 a_t}{3+k_0 a_t} + \frac{1}{9} k_0^2 a_t^2 \right) \label{fric_slip}
\end{equation}

\noindent which reduces to $4 \pi \eta_0 a_t$ for a Newtonian fluid, as expected. In the opposite limit, $k_0\rightarrow \infty$, the boundary condition is not relevant and the friction coefficient is again $V_t \zeta_0 /2$.

We end this section by discussing a few important caveats in the connection between the hydrodynamic theory and the Langevin simulations. While one would directly identify $\zeta_0$ with $n\gamma_0$ in eq. (\ref{Langevin}), possible differences might be relevant in comparisons. On the one hand, the minimum size of the tracer for the hydrodynamic theory to apply is unknown; and on the other hand the compressibility of the colloidal bath (considering only the particles, not the solvent), might be relevant, as the density is diffusive in Langevin systems. Even more, the correct boundary condition on the tracer surface is unknown. 

\section{Results and discussions}
\label{results}

In this section we first test the theoretical results of the modified NSE with simulations, and then analyze the dynamics of a large forced tracer in a bath of colloidal particles. 

\subsection{Friction coefficient of the tracer}

\begin{figure}
\psfig{file=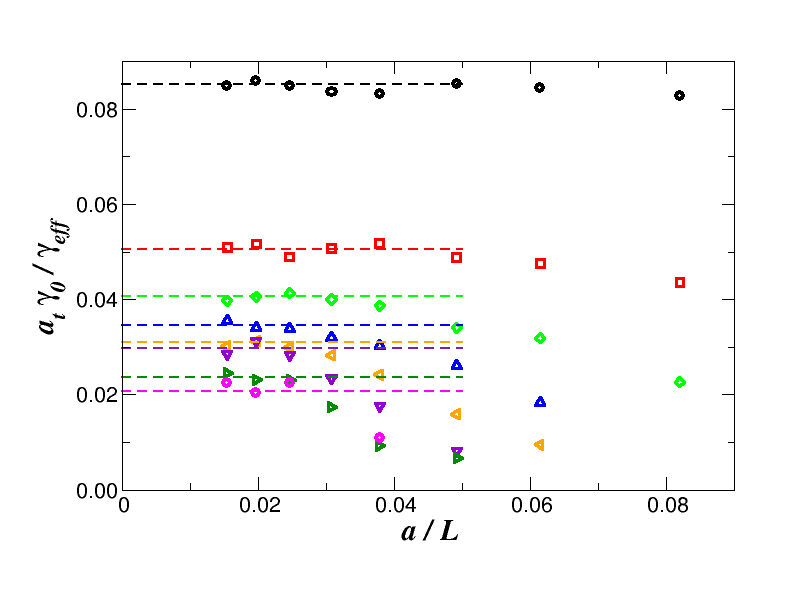,width=0.95\figurewidth}
\caption{Inverse friction coefficient as function of the inverse simulation box size for different tracer sizes (different colors and symbols). From top to bottom: $a_t=1a$, $2a$, $3a$, $4a$, $5a$, $6a$, $7a$ and $8a$.  \label{FSE}}
\end{figure}

The friction coefficient is determined from the steady state tracer velocity, but due to long-range spatial correlations in the bath, it may show importat finite size effects. Figure \ref{FSE} analyses this effect by showing the inverse effective friction coefficient as a function of the inverse box size for different tracer sizes. This representation is motivated by the theoretical analysis of the finite size effects in a Newtonian solvent within the NSE. Hasimoto \cite{Hasimoto1959} showed that the friction coefficient, $\gammaeff$, experienced by an array of tracers follows: 

\begin{equation}
\frac{1}{\gammaeff} = \frac{1}{\gamma_{\infty}} \left(1 - \frac{\cal C}{L}\right) \label{Hasimoto}
\end{equation}

\noindent where $\gamma_{\infty}$ is the friction coefficient measured in an infinite system, $\cal C$ is a constant that depends on the array structure (simple cubic, BCC, FCC, ...) and $L$ is the lattice spacing, namely, the simulation box size. For the simple cubic array, that corresponds to the periodic boundary conditions, ${\cal C}=2.8373\: a_t$ \cite{Hasimoto1959}. Previous simulations of the diffusion of a tracer in a bath of particles, with microscopic Newtonian dynamics, have shown the validity of this result \cite{Yeh2004,Sokolovskii2006}. Furthermore, the value of the friction coefficient extrapolated for the bulk, agrees with the Stokes value using the viscosity (calculated with the Green-Kubo integration of the stress autocorrelation function, as discussed below), and slip boundary conditions.

The data in Fig. \ref{FSE} shows that $\gammaeff^{-1}$ grows for increasing system sizes for small and intermediate $L$, but levels off for large systems. These results clearly deviate from the prediction for a Newtonian fluid, eq. (\ref{Hasimoto}), as expected for Langevin systems with a dissipative term. Notably, it also indicates that the bulk value can be obtained from simulations of large enough systems. In a previous work, it was shown that this general result does not depend on the particular details of the simulation \cite{Orts2019} (considering the volume of the tracer in the system volume, fixing the center mass of the system, or varying the friction coefficient with the solvent). 

\begin{figure}
\psfig{file=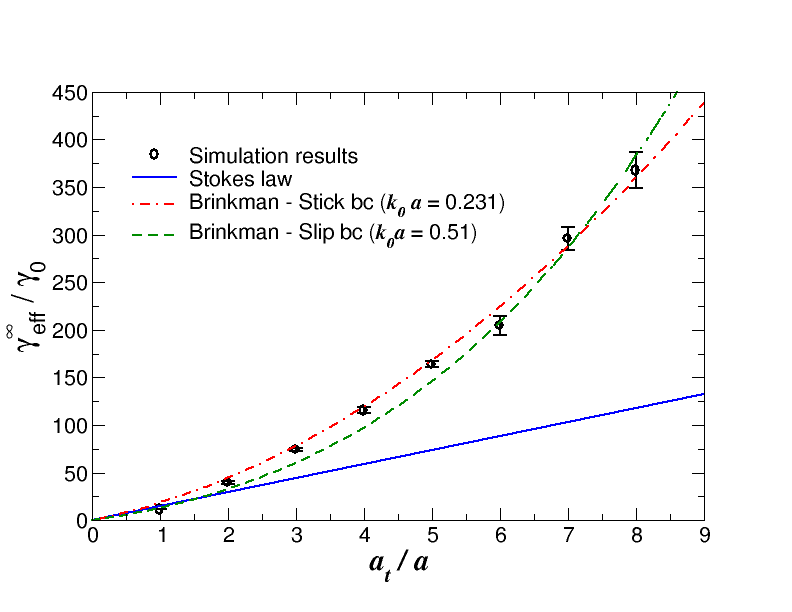,width=0.95\figurewidth}
\caption{Friction coefficient extrapolated to the infinite system as function of the tracer size (the error bars indicate the dispersion of the data for large systems). The lines are the results from Brinkman's equation with stick or slip boundary conditions and the Stokes' law, as labeled. \label{gamma-at}}
\end{figure}

The values of the friction coefficient with an infinite bath, to be compared with the theory, are taken from the plateau for large systems. The results are plotted in Fig. \ref{gamma-at} as a function of the tracer size, with the error bars representing the dispersion of the data. The simulation data deviates clearly from the linear trend predicted by the Stokes' law for a Newtonian fluid, while the Brinkman equation predicts the qualitative behaviour of the friction coefficient adjusting the only unknown parameter $k_0$ (see below). 

To make a more quantitative test of the theoretical models, we calculate the shear viscosity of the bath of quasi-hard particles. This is given by the Green-Kubo relation, namely the integral of the stress autocorrelation function, which accounts for the particle-particle direct interactions as well as the kinetic energy \cite{EvansMorriss}:

\begin{equation}
\eta_0=\frac{\beta}{3V} \int_0^{\infty} dt \sum_{\mu<\nu} \langle \sigma^{\mu\nu}(t) \sigma^{\mu\nu}(0) \rangle
\end{equation}

\noindent where $\beta=1/k_BT$ is the inverse thermal energy, $V$ the system volume and $\sigma^{\mu\nu}(t)$ is the $\mu\nu$-component of the stress tensor. The sum runs over all off-diagonal terms of the stress tensor, and $\langle \sigma^{\mu\nu}(t) \sigma^{\mu\nu}(0) \rangle$ is the stress auto-correlation function. The time integral over the correlation function is more conveniently performed using the Einstein relation \cite{Allen1987,Puertas2007}. 

The Green-Kubo integration gives for the viscosity of the bath $\eta_0=(3.9 \pm 0.1)\,\sqrt{kT m}/a^2$. With this value, the Stokes' prediction is plotted in Fig. \ref{gamma-at} (blue continuous line), which underestimates notably the simulation data for large tracers, although the small size limit is correctly captured. The friction coefficient obtained from the Brinkman equation has been adjusted to reproduce the simulations, using $k_0$ as fitting parameter. The dashed lines in Fig. \ref{gamma-at} show the fittings with the calculations considering stick or slip boundary conditions (red or green lines, respectively). Both fittings are equally acceptable, but they give different values of the fitting parameter $k_0$, as shown in the figure.

From the simulation, identifying  $\zeta_0=n\gamma_0$, we expect $k_0 = \sqrt{n\gamma_0/\eta_0} = 0.39/a$, which is within the range of values provided by both fittings. A small value of $k_0$ corresponds to a system controlled by the viscosity of the bath of particles, as expected due to the high density of the bath (recall that the volume fraction is $\phi=0.50$).

\begin{figure}
\psfig{file=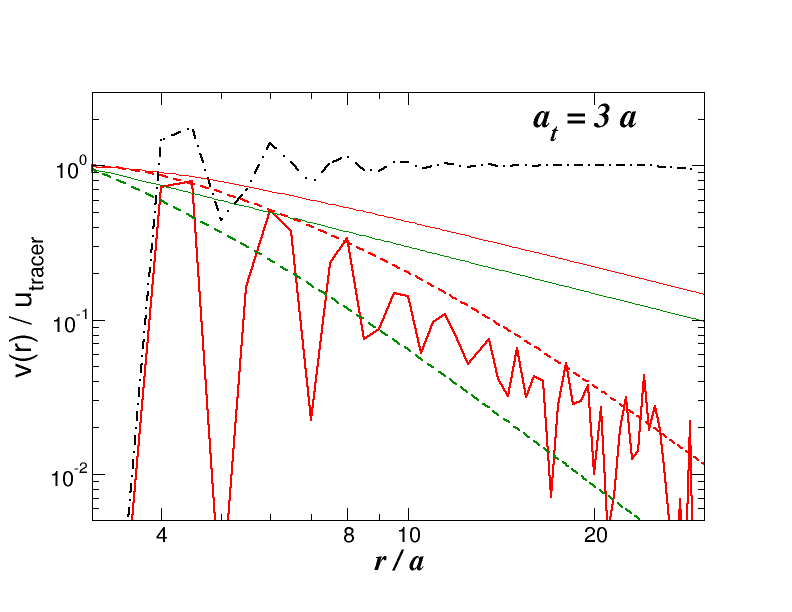,width=0.95\figurewidth}
\psfig{file=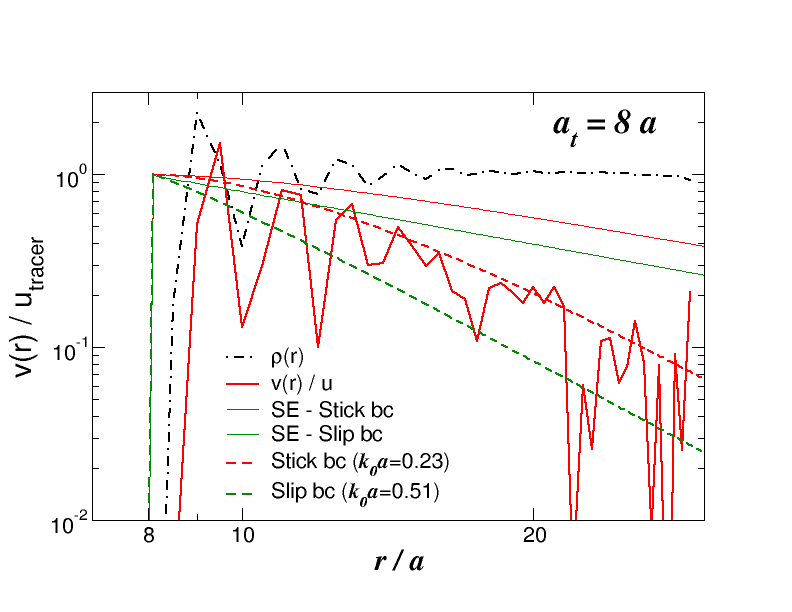,width=0.95\figurewidth}
\caption{Velocity profile in the colloidal bath in front of the tracer from simulations (continuous red line), for two tracer radii, as labeled. 
Theory results for a Newtonian fluid (thin red and green lines) and the Brinkman equation with stick or slip boundary conditions (dashed red and green lines, re.spectively) are also included. 
The dash-dotted black line represents the density of bath particles around the tracer.\label{v-profile}}
\end{figure}

To further compare the model and the simulations, we study the velocity profile in the bath. Fig. \ref{v-profile} shows the velocity of the bath particles in front of the tracer for two tracer sizes and the system with $N=15625$ particles (only the radial component is studied). The distribution of bath particles surrounding the tracer, $\rho(r)$, is also included in the figure to facilitate the interpretation. The velocity profile oscillates in phase with the bath density, and decays faster than the inverse distance, the prediction for the Newtonian fluid, irrespective of the boundary condition. Brinkman's model, eq. (\ref{u-brinkman}), on the other hand, reproduces quite well the decay of the velocity profile (as $1/r^3$), but also quantitatively with the values of $k_0$ obtained from the fitting of the friction coefficient for both boundary conditions, and for both tracer sizes. However, the theory based on the NS equation fails to capture the oscillations due to the finite size of the bath particles, as expected for a continuum model for the bath. Again, both boundary conditions compare equally well with the simulations, bracketing the simulation results.

\begin{figure}
\psfig{file=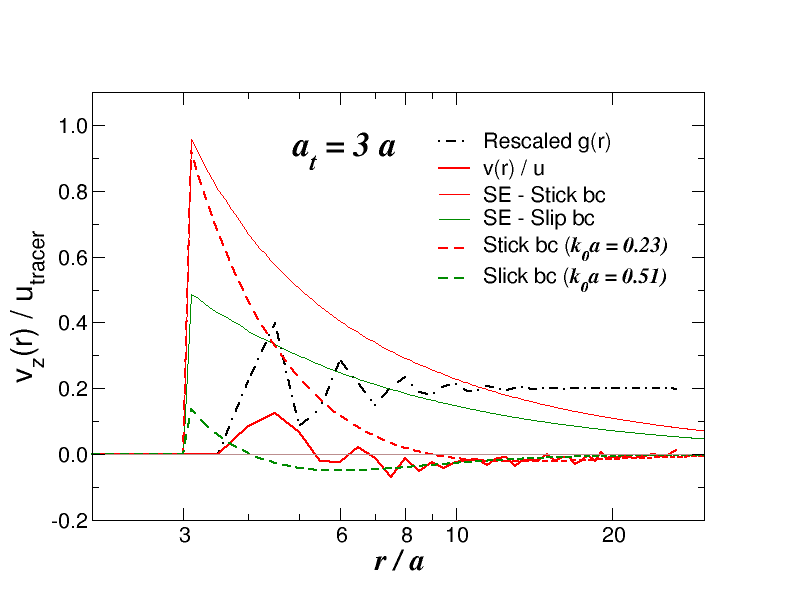,width=0.95\figurewidth}
\psfig{file=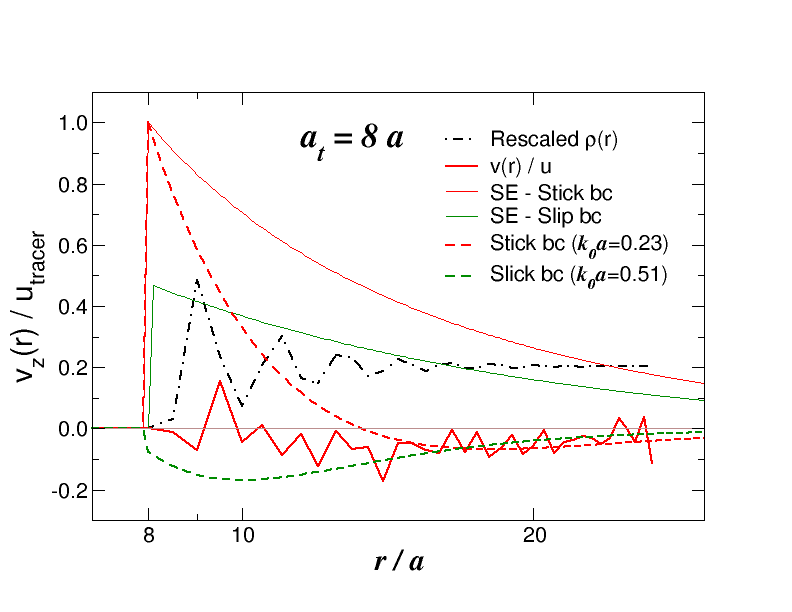,width=0.95\figurewidth}
\caption{$z$-component of the velocity in the colloidal bath in the plane perpendicular to the tracer for $a_t=3a$ (upper panel) and $a_t=8a$ (lower panel). 
Simulations (continuous red line), and theory results for a Newtonian fluid (thin red and green lines) and the Brinkman equation with stick or slip boundary conditions (dashed red and green lines, respectively) are shown. 
The dash-dotted black line represents the density of bath particles around the tracer rescaled to fit into the same scale.\label{v-trans}}
\end{figure}

A more prominent difference between the stick and slip boundary conditions is obtained if the angular component of the velocity field in the direction perpendicular to the external force is studied. This is tackled in Fig. \ref{v-trans} for the same  tracer sizes (the $z$-component of the velocity, parallel to the force, is studied). For small distances from the tracer, the stick boundary conditions result in a positive velocity, which becomes negative further away, but the slip boundary condition produces a negative velocity for all distances. The simulation results agree with both cases for long distances (negative velocity), but are close to zero near the tracer. This result, in conjuction with all previous comparisons, probably indicates that a mixed boundary condition is optimal in describing the friction and velocity fields of the tracer in a colloidal bath with the Brinkman equation. For completeness, the predictions from the NSE for stick and slip boundary conditions are shown, indicating that the behaviour observed in the simulations cannot be reproduced. 

\begin{figure}
\psfig{file=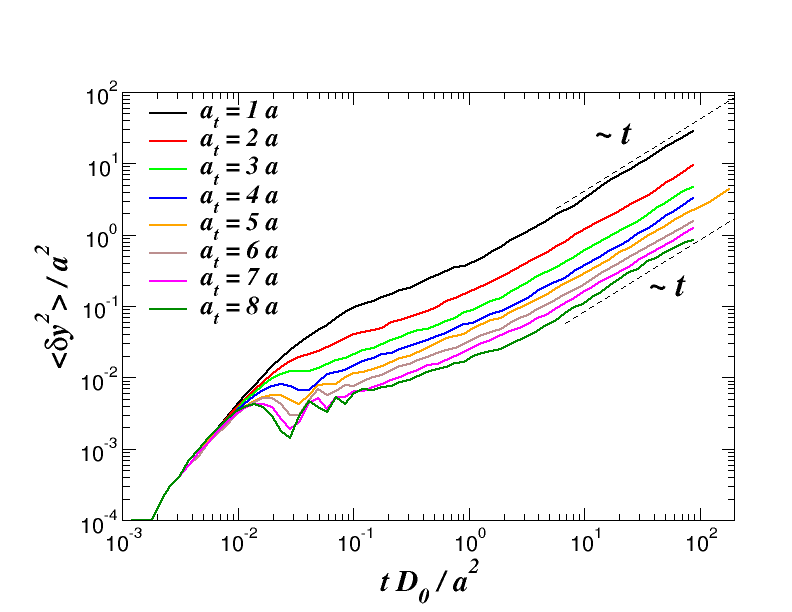,width=0.95\figurewidth}
\psfig{file=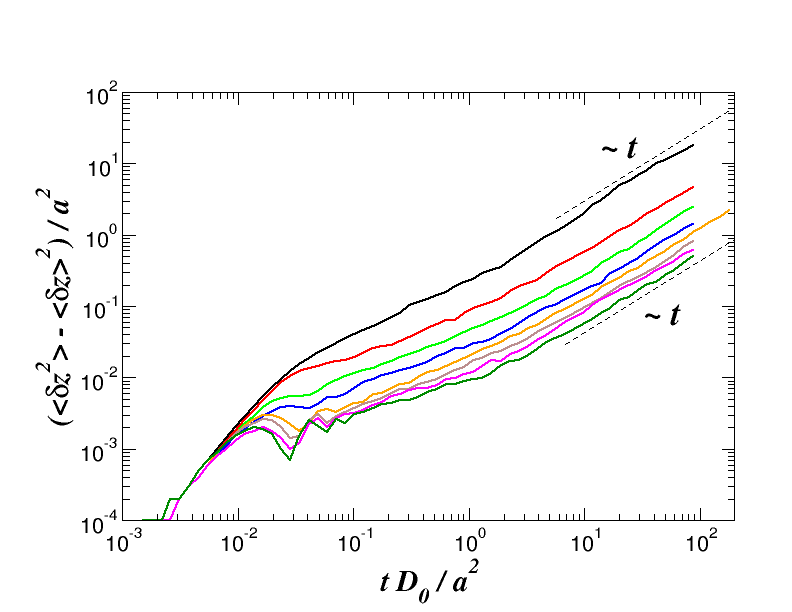,width=0.95\figurewidth}
\caption{Tracer mean squared displacement in the direction perpendicular to the force (upper panel), and parallel to the force (lower panel), for different tracer radii, as labeled
(increasing from top to bottom). \label{msd}}
\end{figure}

\subsection{Tracer dynamics}

In this subsection, we analyze the transient dynamics of the forced tracers of different sizes, for a small force pulling the tracer for $t>0$. Fig. \ref{msd} shows the mean squared displacement of the tracer perpendicular to the force direction and parallel to it (with the drift velocity substracted). Long time diffusion is reached for all tracers, in particular in the longitudinal direction, i.e. superdiffusion is not observed for this density \cite{Winter2012} (superdiffusion has been indeed observed in this same system for larger densities). Notably, the self diffusion coefficient decreases with increasing tracer size, developing a shoulder in the MSD and a sub-linear increase at intermediate times. The typical distance corresponding to the height of the shoulder also decreases with the size of the tracer. Recall that the length unit is the bath particle radius, i.e. if the tracer radius is used, the decrease in the localization length is enlarged, pushing to a tiny fraction of the tracer radius (smaller than $10^{-4}a_t^2$ for the biggest tracer).

\begin{figure}
\psfig{file=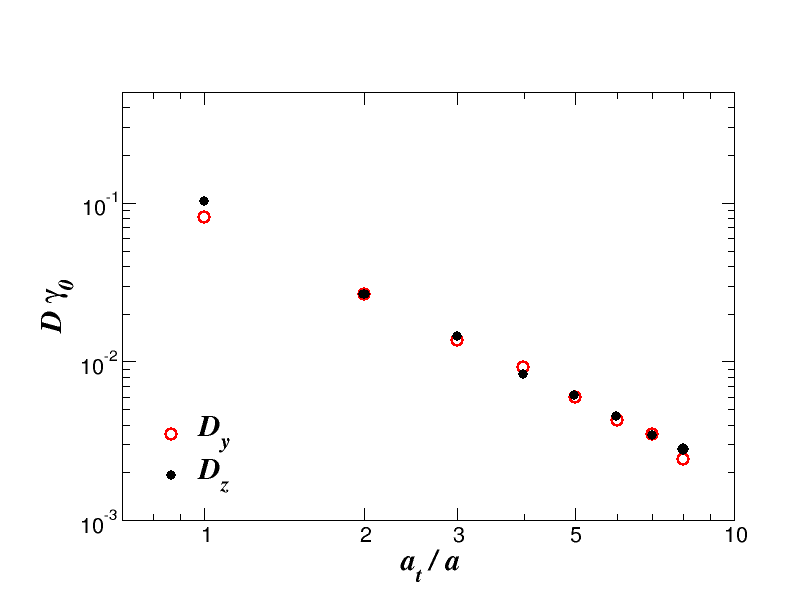,width=0.95\figurewidth}
\caption{Diffusion coefficients in the direction perpendicular and parallel to the external force, as labeled. \label{diff}}
\end{figure}

The self-diffusion coefficients, obtained from the long-time slope of the MSD in both directions, are shown in Fig. \ref{diff}. Both of them are very similar and follow the same trend, decaying almost two decades in the range of tracer sizes studied here. Indeed, not only the slopes of the MSD in both directions are close to each other, but the MSD themselves are very similar (the relative differences are below 20\% in all cases, and constant within the statistical noise). The equality of the MSD in both directions, and the concomitant diffusion coefficients, despite the anisotropy induced by the external force, indicates that the force is small enough to keep the system in the linear regime. 

\begin{figure}
\psfig{file=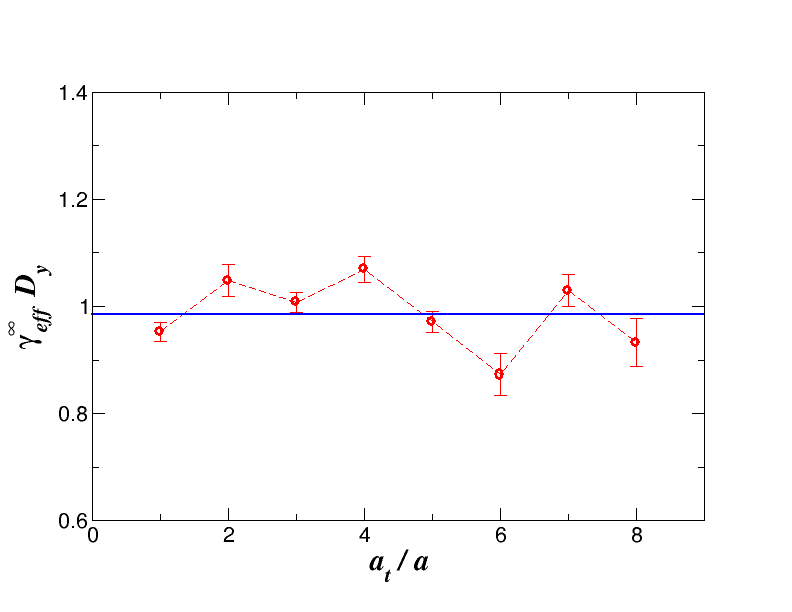,width=0.95\figurewidth}
\caption{Diffusion coefficient in the force direction times the friction coefficient. The blue line is the average over all data\label{dgamma}}
\end{figure}

Finally, we check the Stokes-Einstein relation for the tracer by plotting the product of the diffusion coefficient times the friction coefficient for all tracer radii. Fig. \ref{dgamma} shows these results as a function of the tracer size. The product is close to $1$ in all cases, fluctuating around a mean value of $0.986$, confirming the validity of the Stokes-Einstein relation, or stated more generally, of the linear response formalism. The mobility, viz.~the inverse friction coefficient, of a tracer feeling a small force is proportional to the diffusion coefficient of the unforced tracer, and the  prefactor is given by the thermal energy, which is set to unity in the simulations. 

\section{Conclusions}

The dynamics of a large tracer pulled with a small force in a bath of quasi-hard colloidal spheres has been studied with Langevin dynamics simulations, and with continuum mechanics. The force is small enough to keep this out-of-equilibrium system in the linear response regime. The analysis of finite size effects in the simulations has shown that the correlations in the bath, induced by the moving tracer, decay faster than in a Newtonian fluid, and within the simulation box, if the system is large enough. This has allowed the analysis of the microviscosity without futher extrapolation with the theory. The Navier Stokes equation has been modified, adding a term proportional to the fluid velocity, resulting in an equation identical to the Brinkman equation, albeit our interpretation of the terms is different. This two-fluid model provides a length scale, $k_0^{-1}$, for the crossover from diffusive transverse momentum transport to friction with the solvent, which depends on the viscosities of the two fluids. The resulting friction coefficient for the tracer grows faster than linear, with both stick and slip boundary conditions,   and the velocity profile decays as $\sim 1/r^3$, for finite $k_0$. The results for a Newtonian fluid are recovered in the limit $k_0\rightarrow 0$.

The comparison of the simulations and theory gives semi-quantitative agreement. Fitting $k_0$, the simulation data can be reproduced with the model, both the friction coefficient and velocity profile in the bath for long distances. The value of $k_0$ also  corresponds to the expectation based on the viscosity calculated from the Green-Kubo relation and the solvent friction coefficient. The two-fluid model describes satisfactorily the physical phenomena in colloidal microrheology, and shows that a correct interpretation of the results  requires accounting for colloidal bath particles and solvent. Also, our results apparently point to mixed effective boundary conditions between stick and slip.

The fluctuations of the tracer position have been studied to obtain the mean squared displacement in the direction parallel to the force and perpendicular to it. Diffusion is attained in both cases at long times, after a transient trapping with a typical length decreasing for increasing tracer sizes. Because the system is in the linear response regime, the diffusion coefficients in both directions are similar despite the anisotry provoked by the external force. Furthermore, the Stokes-Einstein relation is fulfilled, confirming the validity of linear response.

\section{Acknowledgements}

We are indebted to Prof. Thomas Franosch for many useful discussions on the Brinkman model and its solutions. AMP acknowledges financial support from the Spanish Ministerio de Ciencia and FEDER under project no. PGC2018-101555-B-I00 and from UAL/CECEU/FEDER (UAL18-FQM-B038-A); FO, GO and EMG also appreciate the support from the project RTI2018-095993-B-I00. F Orts is supported by an FPI Fellowship (attached to Project TIN2015-66680-C2-1-R) from the Spanish Ministerio de Ciencia.

\bibliographystyle{apsrev}

\end{document}